\documentclass[11pt,twoside]{article}
\usepackage{asp2010}

\resetcounters

\begin{document}

\title{Fitting the Chandra LETG spectrum of SS Cygni in outburst with model atmosphere spectra}
\author{V.F. Suleimanov$^{1,2}$, C.W. Mauche$^{3}$, R.Ya. Zhuchkov$^2$, and K. Werner$^1$
\affil{$^1$Institute for Astronomy and Astrophysics, Kepler Center for Astro and Particle Physics,
       Eberhard Karls University, Sand 1, 72076 T\"ubingen, Germany}
\affil{$^2$Kazan (Volga region) Federal University, Kremlevskaya str. 18, 42008 Kazan, Russia}
\affil{$^3$Lawrence Livermore National Lab., L-473, 7000 East Ave., Livermore, CA 94550, USA}
}

\begin{abstract}
The {\it Chandra\/} LETG spectrum of SS Cyg in outburst shows broad ($\approx 5$ \AA ) 
spectral features that have been interpreted as a large number of
absorption lines on a blackbody continuum with a temperature of
250 kK \citep{m:04}. It is most probable that this is the spectrum of 
the fast-rotating optically thick boundary layer on the white dwarf surface.
Here we present the results of fitting this spectrum with high 
gravity hot stellar model atmospheres. An extended set of LTE 
model atmospheres with solar chemical composition was computed for 
this purpose. The best fit is obtained with the following parameters:
$T_{\rm eff}=190$ kK, $\log g=6.2$, and 
$N_{\rm H}=8 \cdot 10^{19}$\,cm$^{-2}$. The spectrum of this model
describes the observed spectrum in the 60--125 \AA\ range reasonably well,
but at shorter wavelengths the observed spectrum has much higher flux.
The reasons for this are discussed.   
The derived low surface gravity supports the hypothesis
of the fast rotating boundary layer.
\end{abstract}

\section{Introduction}

SS Cyg is one of the best-studied cataclysmic variable stars and is a prototype of dwarf nova stars 
\citep{war:95}. X-ray radiation of this close 
binary in quiescence is hard and can be described by an optically thin
hot ($kT \approx 20$ keV) plasma with an observed flux $\approx 2 \cdot 10^{-10}$ erg s$^{-1}$ 
cm$^{-2}$.  In  outburst, the hard X-ray flux decreases by a factor of ten, the plasma temperature 
is reduced to $\sim 6$--8 keV, and an additional soft component appears with a blackbody temperature
$\approx 200$--300 kK \citep{cor:80,mcg:04,ish:09}.

A high-resolution spectrum of the soft component was obtained with the {\it Chandra\/} LETG and was
carefully investigated by \cite{m:04}. The spectrum, which looks like the photospheric spectra of super-soft
X-ray sources 
 \citep{rauch:10, vr:12}, is naturally associated with the radiation of the boundary layer (BL) between
the accretion disk and white dwarf (WD) \citep{ps:79,kley:91}, and can be phenomenologically described by a blackbody spectrum with 
  $T \approx 250$ kK and numerous broad
 absorption features of ions of O, Ne, Mg, Si, S, and Fe; the BL luminosity and WD spin were also
evaluated by \cite{m:04}. Here we present our attempt to fit the {\it Chandra\/} LETG spectrum of SS Cyg using
the spectra of hot stellar model atmospheres that are close to the Eddington limit, and to make more accurate estimates of the BL parameters on this basis.  

\section{Model atmospheres}

To model high temperature atmospheres that are close to the Eddington limit, we
used our version of the computer code ATLAS \citep{Kurucz:70},
modified to deal with high temperatures \citep{Ibrag:03,sw:07}.  
We assumed local thermodynamic equilibrium (LTE) and accounted for  the pressure ionization effects 
using the occupation probability formalism \citep{Hum.Mih:88} as  described by \citet{Lanz.Hub:94}. 
We took into account coherent 
electron scattering together with the 
free-free  and bound-free  opacity  
of all ions of the 15 most abundant elements using opacities from \cite{VYa:95}. 
Line blanketing is taken into account using $\sim 25000$ spectral lines from the CHIANTI, Version 3.0,
atomic database \citep{dere:97}. 

\begin{figure}
\begin{center}
 \includegraphics[height=.22\textheight]{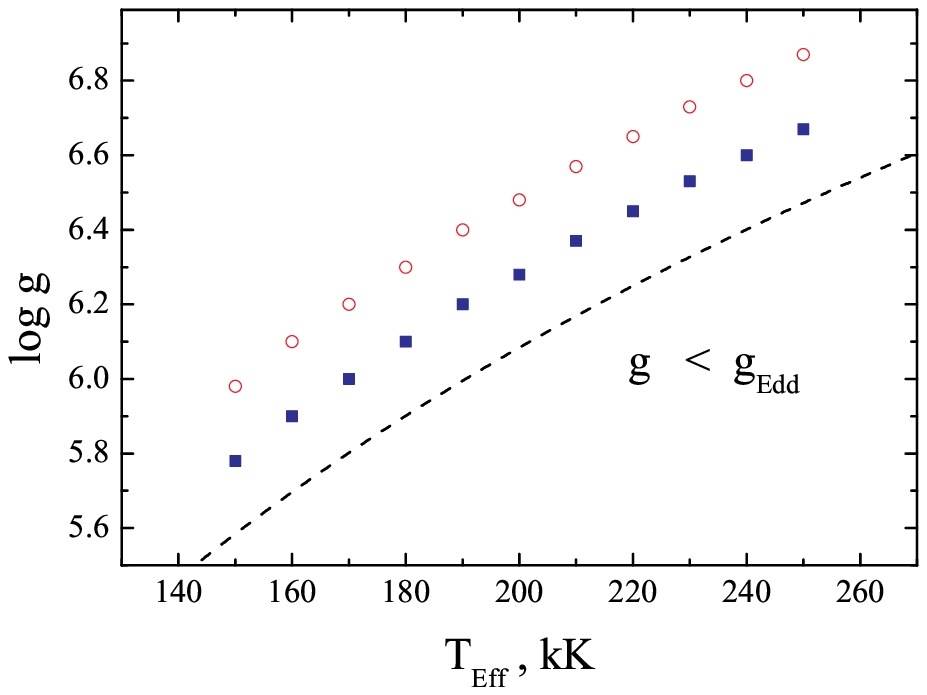}
\includegraphics[height=.23\textheight]{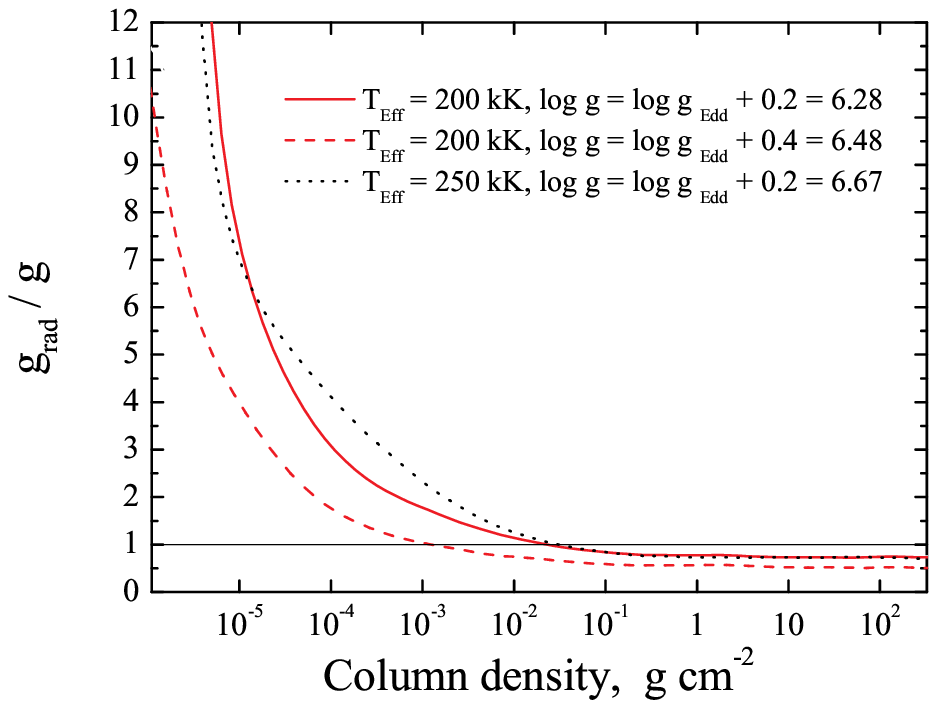}
 \caption{ \label{fig:svt_1}
{\it Left:} Computed model atmospheres on the $T_{\rm eff}$--$\log g$ plane. The locus of points for which
$\log g = \log g_{\rm Edd}$
is shown by the dashed curve. {\it Right:} Ratio of the radiation force to the surface gravity vs.\ depth for various
 model atmospheres.}
\end{center}
\end{figure}

Using our code, we calculated 22 model atmospheres with solar chemical composition. The effective
 temperatures of the models range between 150 kK and 250 kK with a step of 10 kK. We used two values of
 the surface gravity for each effective temperature: $\log g = \log g_{\rm Edd} + 0.2$ and  
 $\log g = \log g_{\rm Edd} + 0.4$, where $\log g_{\rm Edd} = \log (\sigma_{\rm e}\,\sigma_{\rm SB} 
 T^4_{\rm eff}/c) = 4.88 + 4\log (T_{\rm eff}/10^5\,K)$ is the surface gravity that has an equal radiation
  pressure force for a given $T_{\rm eff}$, and $\sigma_{\rm e} \approx 0.34$ g cm$^{-2}$ is the electron 
  scattering opacity for the assumed solar chemical composition. The positions of the computed models on the 
$T_{\rm eff}$--$\log g$ plane are shown in Fig.\,\ref{fig:svt_1} (left panel).  The considered model atmospheres are
very close to the Eddington limit and a radiation pressure force $g_{\rm rad}$
 due to spectral lines  becomes larger than the surface gravity at the upper atmospheric
 layers (see Fig.\,\ref{fig:svt_1}, right panel). We did not consider
moving atmospheres and we took a gas pressure equal to 10\% of the total pressure 
($P_{\rm gas} = 0.1 P_{\rm tot}$) if $g_{\rm rad} > g$ to enforce hydrostatic equilibrium.

\begin{figure}
\includegraphics[height=.43\textheight]{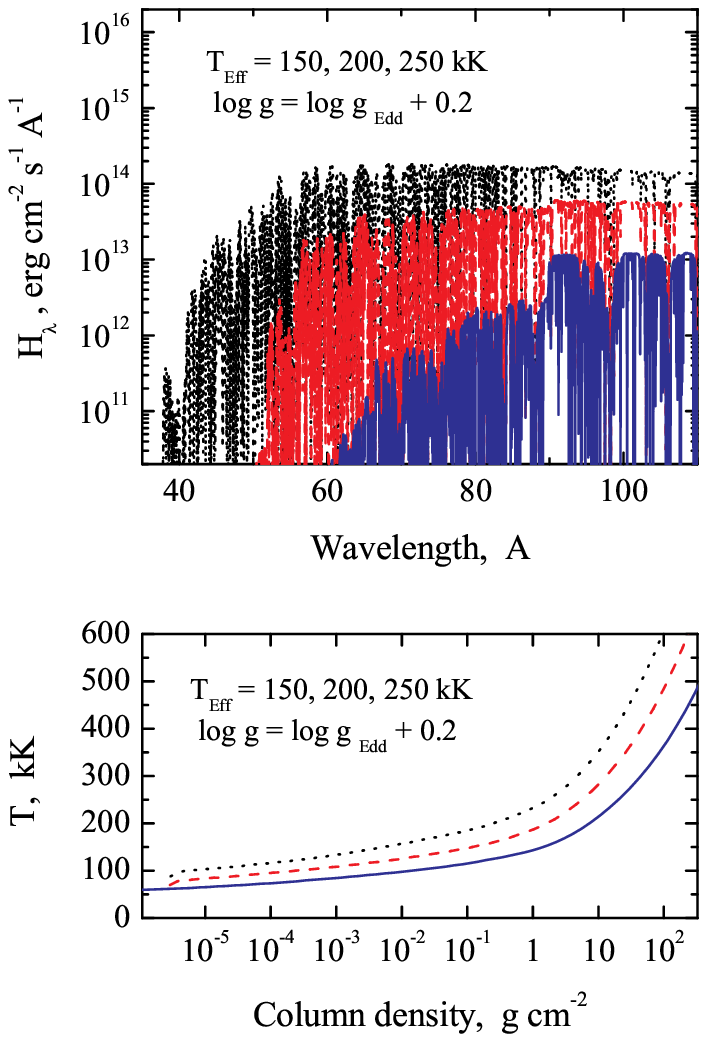}
\includegraphics[height=.43\textheight]{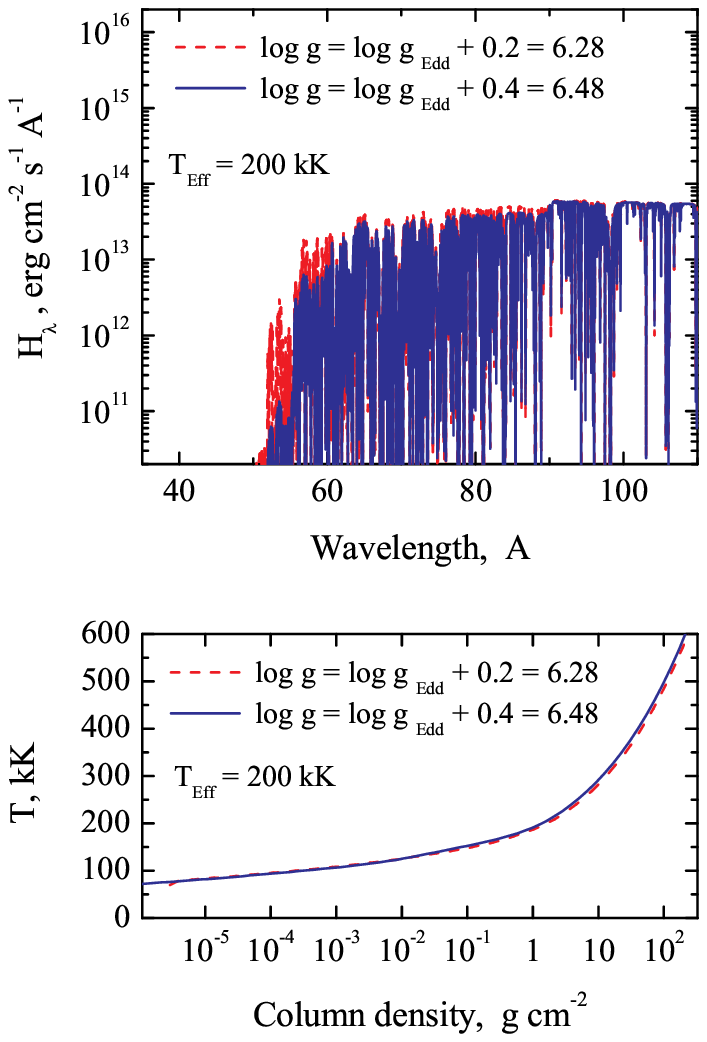}
  \caption{ \label{fig:svt_2}
{\it Left:} Emergent spectra and temperature structures of three model atmospheres with the same differences
 between $\log g$ and the Eddington limit: $\log g = \log g_{\rm Edd} + 0.2$, and various effective temperatures: 
 150 kK (solid curves), 200 kK (dashed curves), and 250 kK (dotted curves). {\it Right:} Emergent spectra 
 and temperature structures of two model atmospheres with the same effective temperature (200 kK) and 
 different $\log g$: $\log g = \log g_{\rm Edd} + 0.2$ (dashed curves) and 
 $\log g = \log g_{\rm Edd} + 0.4$ (solid curves).}
\end{figure}

Examples of the computed emergent spectra together with the temperature structures are shown in 
Fig.\,\ref{fig:svt_2}. The emergent spectra are dominated by absorption line forests and they become
 similar to the observed spectrum of SS Cyg only after accounting for the LETG spectral resolution 
(see Fig.\,\ref{fig:svt_3}). The differences between the spectra of models with different surface gravities
is obvious (see Fig.\,\ref{fig:svt_2}, right panel) and can be found from a comparison with the observed spectrum.


%\clearpage
\section{Results}

\begin{figure}
\begin{center}
\includegraphics[bb= 70 344 570 677,clip,angle=0,height=0.4\textheight]{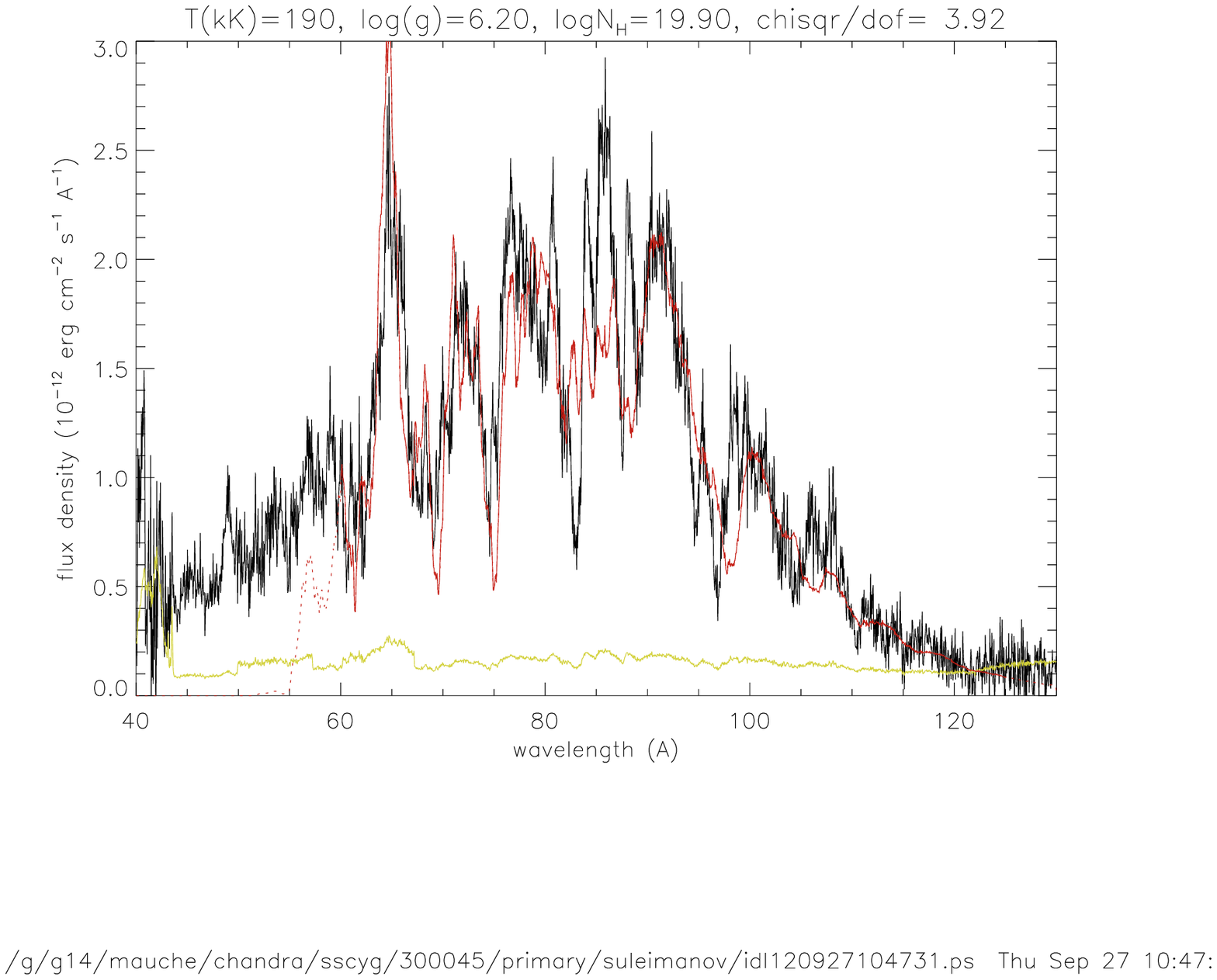}
  \caption{ \label{fig:svt_3}
The {\it Chandra\/} LETG spectrum of SS Cyg in outburst (thick black curve) and the best-fit model 
atmosphere spectrum with $T_{\rm eff} = 190$ kK, $\log g = 6.2$, and $\log N_{\rm H} = 19.9$ 
(thin red curve). The fitting was performed in the 60--125 \AA\ wavelength range. The model spectrum
at the shorter
wavelengths is shown by the dashed curve. 
}
\end{center}
\end{figure}

\begin{figure}
\begin{center}
\includegraphics[height=.32\textheight]{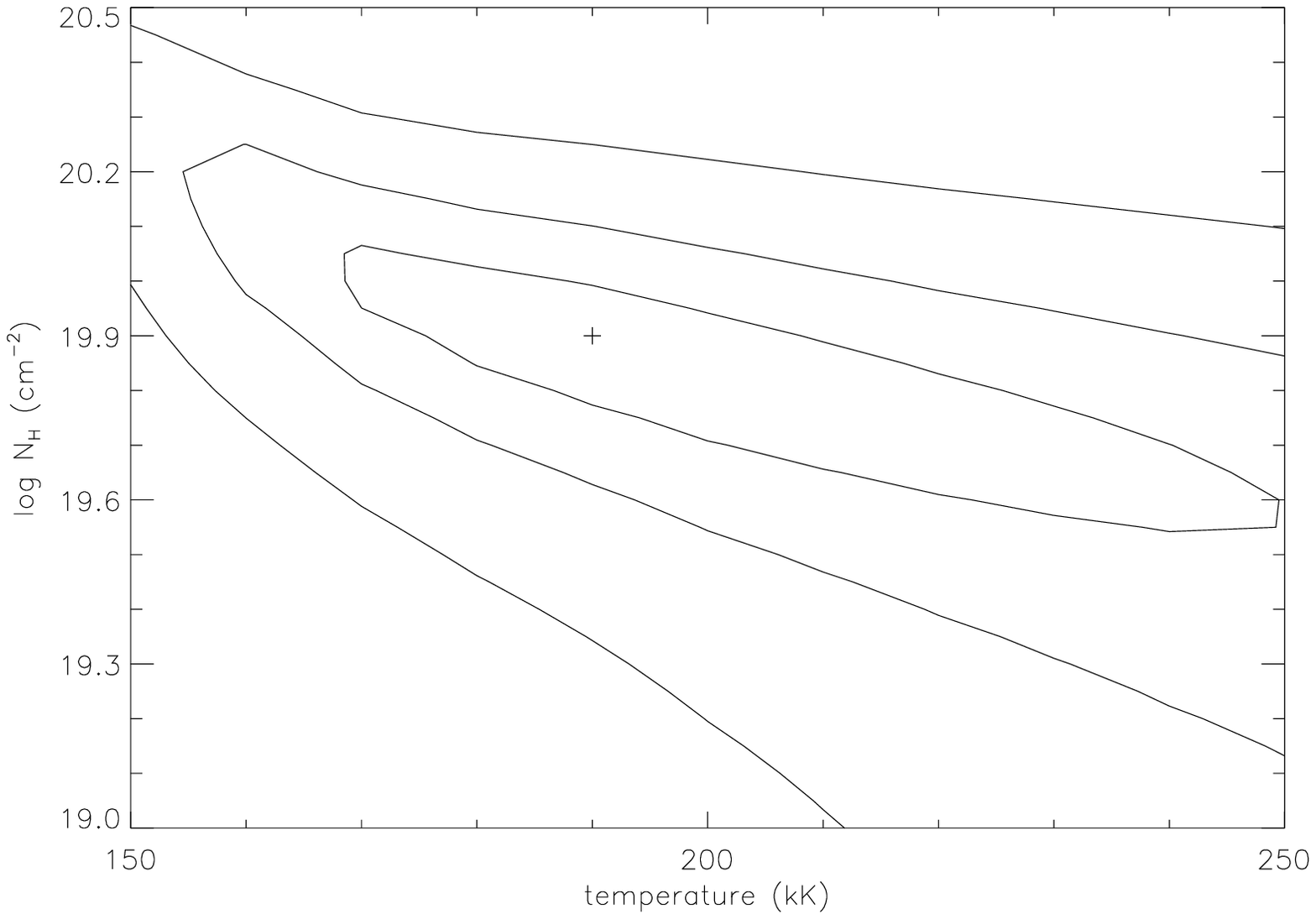}
  \caption{ \label{fig:svt_4}
Position of the best-fit model on the $T_{\rm eff}$--$\log N_{\rm H}$ parameter 
plane and contours of $\chi^2= [1.5,3,6]\, \chi^2_{\rm min}$.
}
\end{center}
\end{figure}

We fit the observed soft X-ray spectrum of SS Cyg using the set of model atmosphere 
spectra described above, convolved with the {\it Chandra\/} LETG
 spectral resolution $\Delta \lambda = 0.05$ \AA . 
The comparison of the best-fit spectrum with the observed spectrum is shown in  Fig.\,\ref{fig:svt_3},
 and the contours of $\chi^2$ on the $T_{\rm eff}$--$\log N_{\rm H}$ parameter plane are shown in 
 Fig.\,\ref{fig:svt_4}. The fitting procedure was performed in the 60--125 \AA \
 wavelength range only,
 because at the shorter wavelengths our model spectra cannot describe the observed spectrum. The possible
 reasons for this are discussed in the next section. We found that the best-fit model parameters  correspond
 to the models with the lower surface gravity 
 $\log g = \log g_{\rm Edd} + 0.2$ with $T_{\rm eff} = 190$ kK, $N_{\rm H} = 8
  \cdot 10^{19}$ cm$^{-2}$, and the normalization $K = f R_{\rm WD}^2/d^2 = 7.82 \cdot 10^{-26}$,
  where $d$ is the distance to SS Cyg and $f$ is the WD
  fractional area occupied by the BL, which can be expressed as the relative BL extension 
along the WD surface $f \approx (2\pi R_{\rm WD}\,2H_{\rm BL})/(4\pi R_{\rm WD}^2) = 
H_{\rm BL}/R_{\rm WD}$.

\begin{figure}
\begin{center}
 \includegraphics[height=.4\textheight]{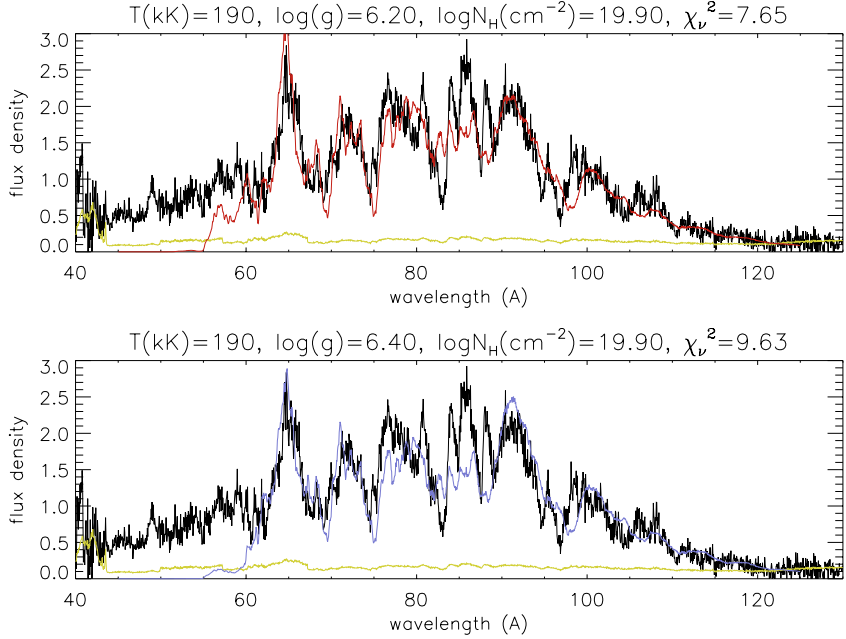}
 \caption{ \label{fig:svt_5}
The {\it Chandra\/} LETG spectrum of SS Cyg in outburst (thick black curve) and best-fit model 
atmosphere spectra with $T_{\rm eff} = 190$ kK, $\log N_{\rm H} = 19.9$, and two surface gravities:
 $\log g = 6.2$ (thin red curve, top panel) and $\log g = 6.4$ (thin blue curve, bottom panel). 
 The fitting was performed in the 45--125 \AA\ wavelength range.
}
\end{center}
\end{figure}

 The obtained fit is not
completely satisfactory (reduced $\chi^2 = 3.9$), hence the formal errors are large and we have not attempted to determine
 them.
However, the choice of the lower surface gravity is statistically justified because the reduced $\chi^2$ is 
significantly larger for fits obtained using the higher gravity models ($\chi^2_{\rm d.o.f.} = 7.65$ vs.\
$\chi^2_{\rm d.o.f.} = 9.63$ for fits performed in the 
45--125 \AA\ wavelength range). This is also clear from Fig.\,\ref{fig:svt_5}, which shows that
 the high gravity spectrum cannot describe the local flux maxima at 78 and 92 \AA .  
 
Using the obtained fit parameters, we evaluated some basic properties of the BL. For ease of comparison, we
 adopted the same system and WD parameters used by \citet{m:04}: 
 $M_{\rm WD} = 1\> M_{\odot}$,  $R_{\rm WD} = 5.5 \cdot 10^8$ cm, 
 $d = 160$ pc, and an accretion disk bolometric luminosity in outburst 
 $L_{\rm Disk} = 10^{35}$ erg s$^{-1}$. In this case, we can evaluate the fractional area of the BL
 $f = 6.3 \cdot 10^{-2}$ ($5 \cdot 10^{-3}$), the bolometric BL
  luminosity  $L_{\rm BL} = 1.8 \cdot 10^{34}$ ($5 \cdot 10^{33}$) erg s$^{-1}$, and the relative BL
   luminosity   $L_{\rm BL}/L_{\rm Disk} = 0.18\, (0.05)$, where
the values obtained by \citet{m:04} are shown in parentheses.  Using the well-known relation between the
 BL and accretion disk luminosities  \citep{K:87,kley:91}  $L_{\rm BL}/L_{\rm Disk} = 
 [1-\Omega_{\rm WD}/\Omega_{\rm K}(R_{\rm WD})]^{2}$, where $\Omega_{\rm K}(R_{\rm WD})$ is the Kepler
 angular velocity at the
 WD radius, we infer that the spin period of the WD in SS Cyg is 12 (9)\,s. 
  
The intrinsic surface gravity of the WD in SS Cyg ($\log g_{\rm WD} = 8.46$) is more than 
two orders higher than
the obtained BL effective surface gravity $\log g_{\rm eff} = 6.2$. 
The surface gravity of the BL can be reduced 
by the fast rotation of the accreting matter, and we evaluated a BL angular velocity $\Omega_{\rm BL} 
\approx 0.98\,\Omega_{\rm K}(R_{\rm WD}) $ using the simple relation 
$g_{\rm eff}  = g_{\rm WD} - \Omega_{\rm BL}^2R_{\rm WD} = g_{\rm WD}\,(1 - 
[\Omega_{\rm BL}/\Omega_{\rm K} (R_{\rm WD})]^2)$. 

\section {Conclusion and Discussion}

On the basis of the above analysis,  we conclude that the BL in SS Cyg can be considered as 
a hot ($\approx 190$ kK), fast rotating [$\Omega_{\rm BL} \approx 0.98\,\Omega_{\rm K}(R_{\rm WD})$],  
narrow  ($H_{\rm BL} \approx 0.063\,R_{\rm WD}$) belt on the WD surface. 

This deduction is founded on the fit of the SS Cyg {\it Chandra\/} LETG spectrum with the model 
atmosphere spectra. The obtained fit is not statistically acceptable ($\chi^2_{\rm d.o.f.} = 3.9$), 
especially at the shorter wavelengths ($< 60$ \AA ) and in the 82--90 \AA\ wavelength region. These deficiencies can be connected with model shortcomings: e.g., the chemical composition may differ from solar, non-LTE effects could be important
 \citep[see, e.g.,][]{rauch:10}, and the atomic data are almost certainly neither complete nor entirely accurate.
 The most important unmodeled effect is atmosphere expansion due to a spectral line driven stellar wind, which can be 
 significant because $g_{\rm rad} > g$ at the outer layers of our model atmospheres  \citep[see also][]{vr:12}.
It is likely that the BL cannot be described by a simple one-zone model and that it has a more complicated 
structure,
with a distribution of effective temperatures and surface gravities over its surface. All of these effects
must be taken into account in further investigations.

\acknowledgements This work is supported by the DFG SFB\,/\,Transregio 7 ``Gravitational Wave
 Astronomy'' (V.S.) and the Russian Foundation for Basic Research (grant  12-02-97006-r-povolzhe-a) (R.Zh.).
 C.W.M.'s contribution to this work was performed under the auspices of the U.S. Department of Energy by Lawrence Livermore National Laboratory under Contract DE-AC52-07NA27344.
   
\bibliography{suleimanov_talk}
\end{document}